\documentclass[10pt, twocolumn]{IEEEtran}

\usepackage[utf8]{inputenc}
\usepackage{todonotes}
\usepackage{mathtools}
\usepackage{cite}
\usepackage{subcaption}
\usepackage{gensymb}
\usepackage{textcomp}
\usepackage{multirow}
\usepackage{amsmath}
\usepackage{amssymb}
\usepackage[linesnumbered, ruled,vlined]{algorithm2e}
\usepackage{array}

\newcolumntype{?}{!{\vrule width 1pt}}

\title{Fast Steerable Wireless Backhaul Reconfiguration}

\author{
\IEEEauthorblockN{Ricardo Santos\IEEEauthorrefmark{1}, Nina Skorin-Kapov\IEEEauthorrefmark{2}, Hakim Ghazzai\IEEEauthorrefmark{3}, and Andreas Kassler\IEEEauthorrefmark{1}\\ \vspace{+0.1cm}
\IEEEauthorblockA{\small \IEEEauthorrefmark{1}Karlstad University, Karlstad, Sweden\\ Email: \{ricardo.santos, andreas.kassler\}@kau.se\\ \vspace{+0.1cm}
\IEEEauthorrefmark{2}University Center of Defense, San Javier Air Force Base, MDE-UPCT, Murcia, Spain\\ Email: nina.skorinkapov@cud.upct.es\\ \vspace{+0.1cm}
\IEEEauthorrefmark{3}Stevens Institute of Technology, Hoboken, NJ, USA,\\ Email: hghazzai@stevens.edu}
{\thanks {\vspace{-0.4cm}\hrule
\vspace{0.1cm} \indent 2019 IEEE. Personal use of this material is permitted. Permission from IEEE must be obtained for all other uses, in any current or future media, including reprinting/republishing this material for advertising or promotional purposes, creating new collective works, for resale or redistribution to servers or lists, or reuse of any copyrighted component of this work in other works.}}
}
}

\begin{document}

\maketitle
\pagestyle{empty}
\thispagestyle{empty}
\pagenumbering{arabic}

\begin{abstract}
Future mobile traffic growth will require 5G cellular networks to densify the deployment of small cell base stations (BS). 
As it is not feasible to form a backhaul (BH) by wiring all BSs to the core network, directional mmWave links can be an attractive solution to form BH links, due to their large available capacity. When small cells are powered on/off or traffic demands change, the BH may require reconfiguration, leading to topology and traffic routing changes. Ideally, such reconfiguration should be seamless and should not impact existing traffic. However, when using highly directional BH antennas which can be dynamically rotated to form new links,  this can become time-consuming, requiring the coordination of BH interface movements, link establishment and traffic routing.
In this paper, we propose greedy-based heuristic algorithms to solve the BH reconfiguration problem in real-time. We numerically compare the proposed algorithms with the optimal solution obtained by solving a mixed integer linear program (MILP) for smaller instances, and with a sub-optimal reduced MILP for larger instances. The obtained results indicate that the greedy-based algorithms achieve good quality solutions with significantly decreased execution time.

\end{abstract}

\section{Introduction}
\label{sec:intro}
Mobile traffic predictions expect the growth of mobile data to reach 49 exabytes per month, by 2021 \cite{cisco2017cisco}. To increase capacity and support new use-cases, 5G networks will focus on the densification of small cell base stations (BS) and improvements on the used wireless spectrum \cite{7456186}. However, the densification of BSs will bring new challenges to the backhaul (BH), as it is not feasible to connect all small cells through fiber-cabled links to the core network.
As an alternative, a 5G wireless BH can be formed by establishing multiple millimeter-wave (mmWave) links, which can provide the required bandwidth to forward the user equipment (UE) traffic. By forming multi-hop paths, the mmWave small cells can forward aggregated UE traffic towards gateway nodes, forming a dense BH mesh topology, where each node can potentially form links with multiple neighbors.

When traffic demands change over time, the wireless BH can be reconfigured by adaptively turning on/off small cells to provide localized capacity on demand. By turning off not needed BH nodes, the BH energy consumption and operational costs can be reduced. Additionally, if any of the BH mmWave links fails due to e.g. long-lasting obstacle blockage, the topology should be adapted to provide forwarding alternatives. Ideally, changes in topology due to new nodes and links being activated or updates of the forwarding states should be seamless to existing UE traffic.

Due to the path loss properties of mmWave links, highly directional antennas with high gain are needed \cite{6515173}. This requires either the sectorization of multiple large antenna arrays (e.g. with 8 $\times$ 8 elements), or using passive reflect arrays/lenses, that focus high gain beams on a single focal point. Such passive reflect arrays can be mounted on mechanical steerable platforms \cite{uchendu2016survey}, where the antennas can be rotated and aligned to form links with different neighbor small cell BH nodes.
When a new link must be formed, its mechanical alignment is not immediate and can take several seconds to be completed\cite{steerable2019}, during which traffic routed over that link is lost. Consequently, a seamless BH reconfiguration becomes challenging.

To manage the wireless BH, software-defined networking (SDN) based approaches have been proposed, where the (re-)configuration is handled by a centralized control plane entity. The SDN control plane is not only responsible for  BH forwarding, but also for interface alignment and configuration of BH links between small cells. While SDN-based wireless BH architectures have been previously deployed in indoor and outdoor testbeds \cite{steerable2019, 7561000, 5gxhauld53}, the calculation of new BH configuration states has only been studied using  mathematical models. 
Examples optimize the BH topology, routing, and UE assignment, while minimizing the energy consumption \cite{tran2015dynamic, mesodiakaki2017joint, ogawa2017traffic}.
However, the orchestration of the different steps to transition from an existing topology to a new one has not been thoroughly considered. 
Such orchestration is an NP-hard problem which involves coordinating the alignment of the mmWave transceivers, establishing new links, and rerouting existing BH traffic. In \cite{8647747}, we proposed an exact mixed integer linear program (MILP) for optimal BH reconfiguration to minimize the total packet loss. Although the proposed MILP optimally orchestrates the reconfiguration, it does not scale due to the problem complexity.

This paper presents fast and scalable greedy-based heuristic algorithms for the reconfiguration of the steerable wireless BH. The main greedy steerable BH reconfiguration algorithm (Greedy-SBRA) selects temporary BH links to be established during the reconfiguration in order to reduce the packet loss. It calculates the required antenna movement to form those links, based on a ranking function that considers different link attributes and their respective weights. To achieve high-quality results, the Greedy-SBRA requires prior parameter tuning to select the best weights for the link attributes, for each problem instance. While this approach is significantly more scalable than the MILP, it can still be time consuming and not entirely suitable for online reconfiguration.
Therefore, we propose a randomized multi-start variant, called \textit{MS Greedy-SBRA}. This variant iteratively runs the Greedy-SBRA with different weights, while randomizing the algorithm's link selection phase, for further solution diversification. 

We evaluate our algorithms with respect to packet loss and algorithm execution time for multiple topologies, maximum reconfiguration time, and available antennas per node. Our numerical results demonstrate that the greedy algorithms can achieve good quality solutions with significantly reduced execution time for the test cases where the optimal MILP could be solved. For larger cases, we compare with a less complex MILP (referred to as PVF-MILP), that provides sub-optimal solutions by fixing a set of decision variables from the MILP problem. For these problem instances, the proposed heuristics in most cases obtain better results in reduced time. 

The paper is structured as follows. Section~\ref{sec:problem} describes the steerable wireless BH problem definition. Section~\ref{sec:algorithms} presents our proposed reconfiguration algorithms. Numerical results are detailed in Section~\ref{sec:evaluation} and the paper is concluded in Section~\ref{sec:conclusions}.

\section{Problem Definition}
\label{sec:problem}
We consider a small cell mesh BH formed by directional mmWave links, composed by a set $\mathcal{D}$ of small cell nodes with $D$ elements. Each BH node $d \in \mathcal{D}$ is located at $pos_d=[x_d, y_d]$.
Additionally, we assume a subset $\mathcal{I}\subseteq\mathcal{D}$ of the small cell nodes is connected to the core network through a fiber link with unlimited capacity. Each node $d$ has a set of wireless network interfaces $\mathcal{N}$ with $N$ elements. Every interface $n$ of each node $d$ is composed by a mmWave transceiver, placed on an independent mechanical rotational platform. Each mechanical platform can rotate horizontally over 360\textdegree~and vertically between -45\textdegree~and 45\textdegree. We assume that all antennas rotate with the same speed and are all calibrated to have the same reference position at 0\textdegree. For simplicity, we focus on the alignment over the horizontal axis.

To form BH links, two interfaces from different nodes within line-of-sight need to be aligned. A binary matrix $\boldsymbol{\delta}$ with $D \times D$ elements defines the nodes that can form mmWave links. A matrix $\boldsymbol{V}$ with same dimensions lists the required angles to align the BH nodes' interfaces, based on the values from $pos_d$. 
We assume the links operate in the 60 GHz band and calculate their average throughput $\boldsymbol{R}_{dd'}$, considering the path loss due to propagation and atmospheric conditions, as in \cite{7523521}. Each BH node serves a given traffic demand $\rho_d$ (measured in Mbps), corresponding to the associated UE requirements. We only consider downstream traffic, as we assume that upstream traffic does not have high bandwidth requirements and can share the links with downstream traffic. Therefore, we create the wireless mesh topology by forming unidirectional links from the core-connected BH nodes to the rest of the BH.

Our objective is to orchestrate the BH reconfiguration to transition from an initial state C1 to a final state C2 as seamlessly as possible, by minimizing the disruption of UE traffic, i.e. the total packet loss until C2 is established. Such reconfiguration between C1 and C2 requires the coordinated realignment of the BH interfaces, forming new links, and rerouting the traffic, given a limited amount of time $T_K$ (in seconds). A proper reconfiguration can be orchestrated by a SDN control plane, where the SDN controller or a dedicated computational entity calculates and triggers the proper antenna re-alignment to form temporary links that are used to establish backup paths while new links are being formed \cite{steerable2019}. If the reconfiguration is not properly orchestrated, existing links can break before backup  paths are available, leading to high packet loss. When the traffic demand changes significantly, state C2 can be different from C1, requiring new links to be formed and routes to be updated in order to avoid congestion.

\begin{figure}[t]%
\centering
\hspace*{\fill}
\begin{subfigure}{.40\columnwidth}
\includegraphics[width=\columnwidth]{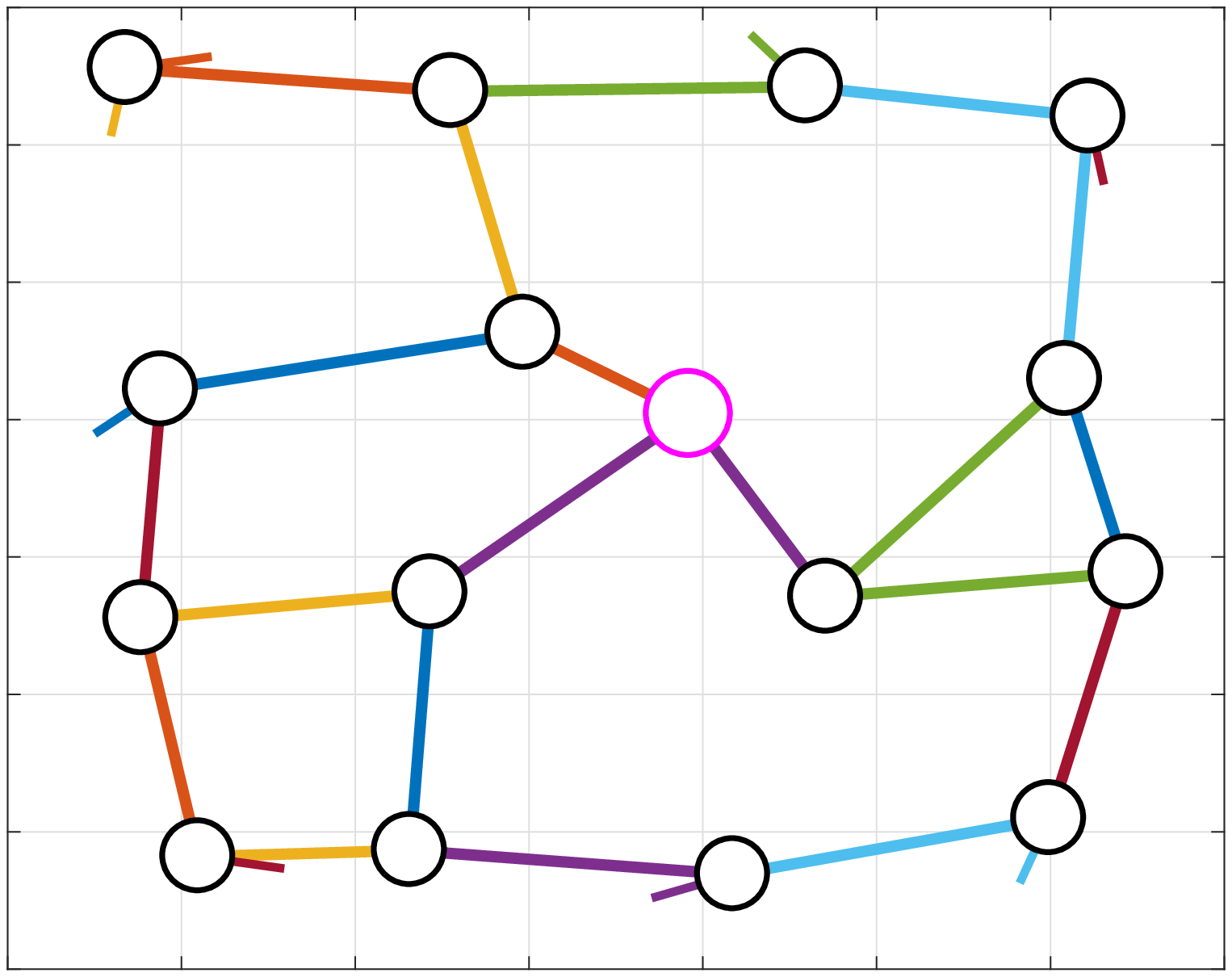}
\caption{Initial state ($k=1$)}%
\label{subfig:example1}%
\end{subfigure}\hfill%
\begin{subfigure}{.40\columnwidth}
\includegraphics[width=\columnwidth]{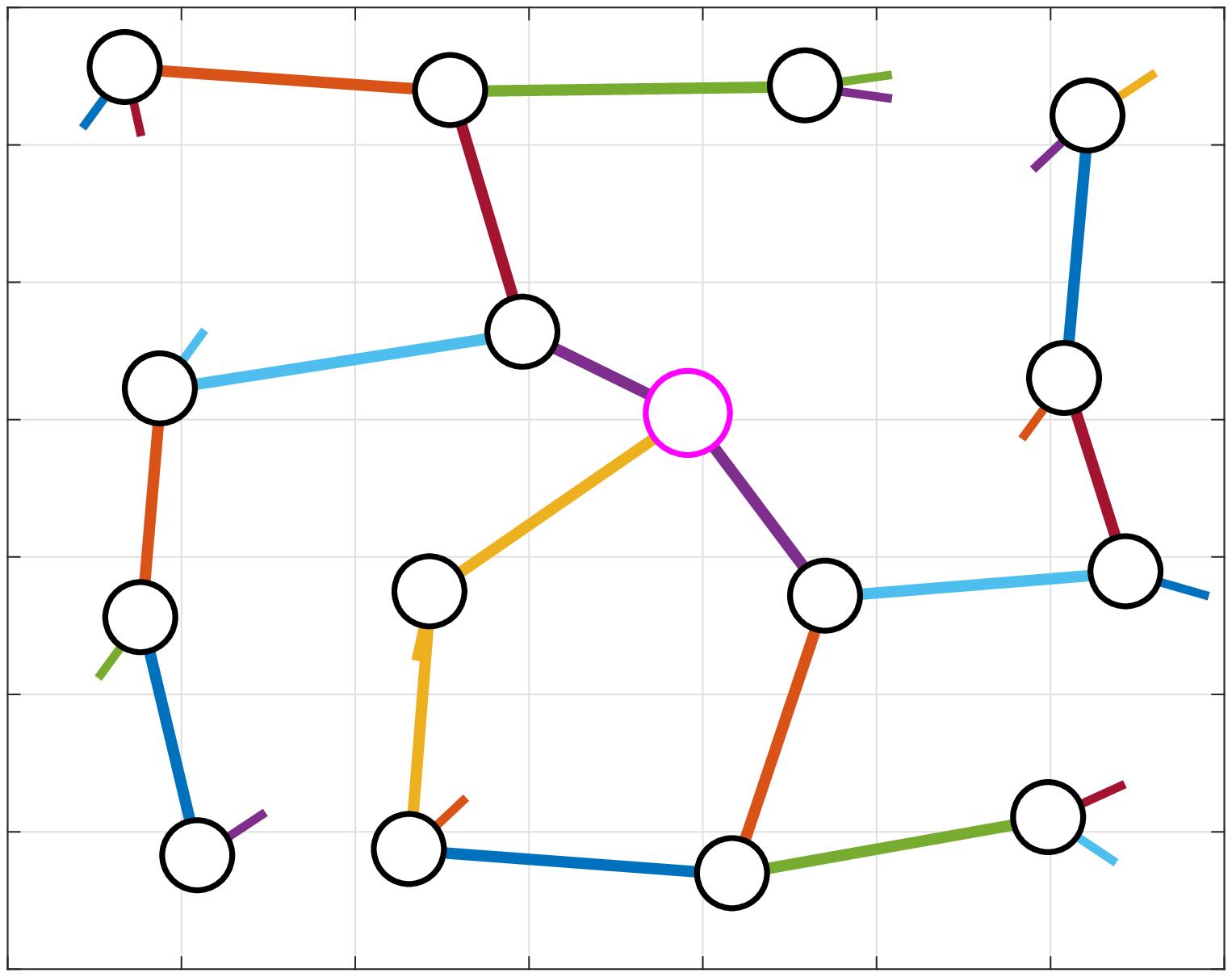}
\caption{Final state ($k=K$)}%
\label{subfig:example2}%%
\end{subfigure}
\hspace*{\fill}
\caption{Example of different backhaul configuration states.}
\label{fig:example-topology}
\end{figure}

State C1 is defined by the initial alignment of all the antennas, given by a matrix $\boldsymbol{A}^0$ with $D \times N$ elements. In addition, a set $\mathcal{X}_{init}$ lists the initially formed links, where each link $(d_nd'_{n'}) \in \mathcal{X}_{init}$ corresponds to interface $n$ of node $d$ connected to interface $n'$ of node $d'$. Similarly, C2 is defined by a set of links $\mathcal{X}_{end}$ with the final (target) topology links. We assume time $T_K$ is divided into $K$ time slots, and each time slot $k$, where $k=1,\dots, K$, has a duration of $\tau$ seconds. We define $\tau$ as the time required to rotate an interface by $\theta$\textdegree, i.e. the minimum rotation angle. As an example, in Figure~\ref{fig:example-topology}, the initial C1 state is given at $k=1$ (Figure~\ref{subfig:example1}), while the final C2 state is reached at $k=K$ (Figure~\ref{subfig:example2}).

% \begin{figure}[t]%
% \centering
% \begin{subfigure}{.48\columnwidth}
% \includegraphics[width=\columnwidth]{figs/shift_D16_N3_1_plot}
% \caption{Initial state ($k=1$)}%
% \label{subfig:example1}%
% \end{subfigure} \hfill%
% \begin{subfigure}{.48\columnwidth}
% \includegraphics[width=\columnwidth]{figs/shift_D16_N3_2_plot}
% \caption{Final state ($k=K$)}%
% \label{subfig:example2}%%
% \end{subfigure}
% \caption{Example of different backhaul configuration states.}
% \label{fig:example-topology}
% \end{figure}

\section{Greedy-based Heuristic Algorithms}
\label{sec:algorithms}
In \cite{8647747}, we proposed an exact MILP for the described steerable wireless BH reconfiguration problem. Since the problem is NP-hard, herein we propose scalable greedy-based heuristics which can solve larger instances in reduced time.

\subsection{Greedy Steerable Backhaul Reconfiguration Algorithm (Greedy-SBRA)}
\label{subsec:greedy}

\begin{algorithm}[t]
\SetAlgoLined
\KwIn{$D$, $N$, $K$, $\theta$, $\mathcal{X}_{init}$, $\mathcal{X}_{end}$, $\boldsymbol{V}$, $\boldsymbol{A}^0$, $\boldsymbol{\delta}$, $\boldsymbol{R}$, $\rho$, $\mathcal{I}$, $\mathcal{W}$}
\KwOut{$CW$, $CCW$, \textit{TotalLoss}}

\tcc{Pre-processing phase}
  Calculate required time slots to rotate each interface to other nodes\;
  Calculate minimum time slots to form each link\;
  Calculate maximum active link time (MALT)\;
  \textit{PossibleLinks} $\gets$ All links that can be active (MALT $> 0$)\;

\textit{LinkScores} $\gets \emptyset$; \textit{FinalLinks} $\gets \emptyset$; \textit{FinalIfaces} $\gets \emptyset$\;

\tcc{Link ranking phase}
\ForEach{$(d_nd'_{n'}) \in$ PossibleLinks}{
  $score_{(dnd'n')} \gets$ GetAttributes$\left((d_nd'_{n'})\right)$ $\times$ $\mathcal{W}$\;
  \textit{LinkScores} $\gets$ \textit{LinkScores} $\cup$ $score_{(dnd'n')}$\;
}

\textbf{sort} \textit{LinkScores} by decreasing order\;

\tcc{Final link selection phase}
\While{LinkScores $\neq \emptyset$}{
  $(d_nd'_{n'}) \gets$ Extract link from \textit{LinkScores}\;

  Add $(d_nd'_{n'})$ to \textit{FinalLinks}\;
  Add $d_n$ and $d'_{n'}$ to \textit{FinalIfaces}\;
  
  Remove all links from \textit{LinkScores} with $d_n$ and $d'_{n'}$\;
}

\tcc{Assignment of movement decision variables}
\ForEach{$(d_nd'_{n'}) \in$ FinalLinks}{
  Update $CW$ and $CCW$, for $d_n$ and $d'_{n'}$\;

  \If{$d_n \in \mathcal{X}_{end}$ }{
    $(d_nd''_{n''}) \gets$ Link from $\mathcal{X}_{end}$ with $d_n$\;

    \If{$d''_{n''} \notin$ FinalIfaces}{
      Update $CW$ and $CCW$, for $d''_{n''}$\;
    }
  }

  \If{$d'_{n'} \in \mathcal{X}_{end}$ }{
    $(d'_{n'}d''_{n''}) \gets$ Link from $\mathcal{X}_{end}$ with $d'_{n'}$\;

    \If{$d''_{n''} \notin$ FinalIfaces}{
      Update $CW$ and $CCW$, for $d''_{n''}$\;
    }
  }
}

\tcc{Post-processing and routing phase}
Build a topology for all $K$ time slots\;
\textit{TotalLoss} $\gets$ LP routing for all $K$ topologies\;

\caption{Greedy-SBRA pseudo-code}
\label{alg:greedy-algorthm}
\end{algorithm}

\begin{table}
\begin{center}
\caption{Input parameters and variables description.}
\label{tab:input-parameters}	
\begin{tabular}{ ll}
\hline
\textbf{Notation}						& \textbf{Description} \\
\hline
\multicolumn{2}{c}{\textit{Greedy-SBRA input parameters }} \\
\hline					 					
$D$			            & Number of BH mesh nodes \\
$N$			            & Number of wireless interfaces per node \\
$K$						& Number of reconfiguration time slots \\
$\theta$				& Interface rotation angle per time slot \\
$\mathcal{X}_{init}$	& Links established at initial state \\
$\mathcal{X}_{end}$		& Links established at final state \\
$\boldsymbol{V}$		& BH nodes' alignment angles \\
$\boldsymbol{A}^0$      & Initial interface alignment values \\
$\boldsymbol{\delta}$	& Possible BH links \\
$\boldsymbol{R}$		& Average BH link throughput \\
$\rho$				    & BH traffic demands \\
$\mathcal{I}$			& Nodes connected to the core network \\
$\mathcal{W}$			& Input link attribute weight set \\

\hline
\multicolumn{2}{c}{\textit{Greedy-SBRA output variables}} \\
\hline
$CW$					& Interface clockwise movement \\
$CCW$					& Interface counter-clockwise movement \\
$TotalLoss$             & Reconfiguration packet loss per node \\

\hline
\multicolumn{2}{c}{\textit{Multi-start Greedy-SBRA parameters}} \\
\hline
$\Omega$				& Number of random weight sets to test \\
$\xi$					& Ranked link list extraction value \\
$I$						& Number of \textit{Greedy-SBRA} iterations \\

\hline
\end{tabular}
\end{center}

\end{table}

The main goal of the Greedy-SBRA is to form links during the reconfiguration from C1 to C2, using links from both states and additional temporary links. During a pre-processing phase, we assign a score to each possible link, based on a set of computed attributes. 
Then, we apply a greedy procedure to iteratively select the most promising links to form. For those links, we fix their interface movement decision variables, maximizing their active duration during the reconfiguration. Once all links are defined over all time slots, we use a linear program (LP) to solve the associated traffic routing, which minimizes the total loss.

The algorithm pseudo-code is given in Algorithm~\ref{alg:greedy-algorthm} and detailed below. Additionally, Table~\ref{tab:input-parameters} lists its respective input parameters and variables.

\subsubsection{Pre-processing phase}
The pre-processing phase determines which links could be possibly formed during the reconfiguration, based on the interfaces' alignment in C1 and links from C2. 
For each link, a set of attributes is calculated, which is used to evaluate the link's potential in the subsequent algorithm phases.
Initially, we compute the minimum number of time slots needed to rotate each interface $n$ from each node $d$ ($d_n$) to every possible neighbor $d'$ (line 1). We calculate the clock and counter-clockwise rotation distances from the initial position $\boldsymbol{A}^0_{dn}$ to the destination $d'$ (given by $\boldsymbol{V}_{dd'}$), dividing it by $\theta$. 
The minimum number of time slots to form each link $(d_nd'_{n'})$  is given as the maximum value between the rotation of $d_n$ to $d'$, and $d'_{n'}$ to $d$ (line 2).
In line 3, we calculate the maximum active link time (MALT) for each link, which is the remaining time slots after the link is formed.
If \textit{a)} the link is part of $\mathcal{X}_{end}$, or \textit{b)} both interfaces $d_n$ and $d'_{n'}$ do not belong to any link from $\mathcal{X}_{end}$, we stop processing the link.
Otherwise, one or both interfaces from ($d_nd'_{n'}$) belong to a different link from $\mathcal{X}_{end}$, and we verify if there are enough time slots to establish the corresponding final link(s), after $(d_nd'_{n'})$ is set.
If this transition is possible on both interfaces, the MALT of $(d_nd'_{n'})$ is subtracted by the maximum number of time slots required to transition to the final link, between the two interfaces. 
If it is not possible to reach the final links at $k=K$, the MALT is set to $0$. Otherwise, the link is added to the set of possible links that can be formed (line 4).

\begin{figure}[t]%
\centering
\hspace*{\fill}
\begin{subfigure}{.27\columnwidth}
\includegraphics[width=\columnwidth]{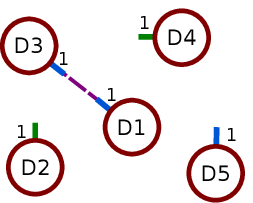}%
\caption{Initial topology}%
\label{subfig:ex-init}%
\end{subfigure} \hfill%
\begin{subfigure}{.27\columnwidth}
\includegraphics[width=\columnwidth]{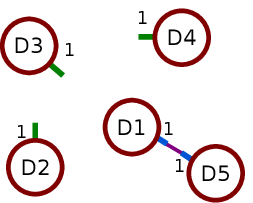}%
\caption{Final topology}%
\label{subfig:ex-end}%
\end{subfigure}
\hspace*{\fill} \\
\begin{subfigure}{.27\columnwidth}
\includegraphics[width=\columnwidth]{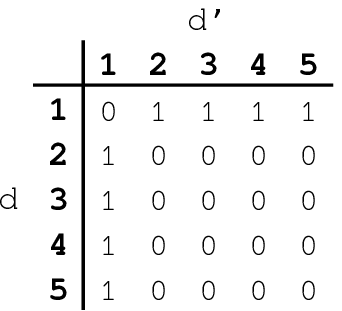}%
\caption{$\boldsymbol{\delta}$ values}%
\label{subfig:delta-values}%
\end{subfigure} \hfill
\begin{subfigure}{.60\columnwidth}
\includegraphics[width=\columnwidth]{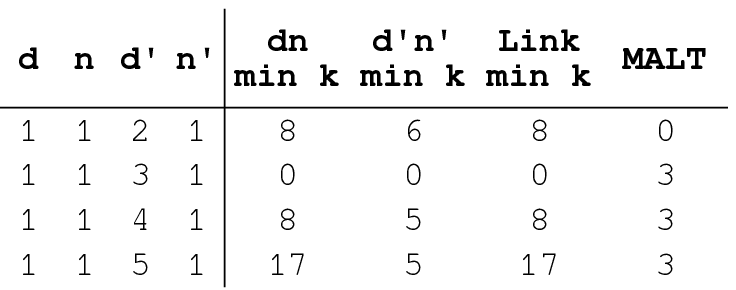}%
\caption{Pre-processing values}%
\label{subfig:table-values}%
\end{subfigure} \hfill
\caption{Example of the Greedy-SBRA pre-processing phase.}
\label{fig:greedy-example}
\end{figure}

Figure~\ref{fig:greedy-example} illustrates this phase for $D=5$ and $N=1$. Given the $\boldsymbol{\delta}$ matrix in Figure~\ref{subfig:delta-values} (with limited possible links, for simplicity), the initial topology in Figure~\ref{subfig:ex-init}, and the final topology in Figure~\ref{subfig:ex-end}, the table in Figure~\ref{subfig:table-values} shows the pre-processed link attributes (lines 1--3).
In this example, $\mathcal{X}_{init}$ is formed by $\{(1_13_1)\}$ and $\mathcal{X}_{end}$ by $\{(1_15_1)\}$, and we assume $K=20$ and $\theta=10^{\circ}$.
Considering the minimum time slots to rotate each interface from link $(1_15_1)$, $1_1$ needs to do a $\approx 170^{\circ}$ clockwise rotation towards node $5$, therefore this value is set to $\lceil\frac{170}{10}\rceil = 17$ (with  interface $1_1$ on link $(1_13_1)$, this value is $0$, as it is already aligned at $k=1$).
The minimum number of slots to form the $(1_15_1)$ link is set to $17$, as interface $1$ from node $5$ only needs $5$ slots to perform its $\approx 50^{\circ}$ counter-clockwise rotation.

Because link $(1_15_1)$ is part of $\mathcal{X}_{end}$ and it needs $17$ slots to be formed, its MALT is $20-17=3$. Since $1_1$ needs to form link $(1_15_1)$ at $K=20$, the MALT of link $(1_14_1)$ is $20-8-(17-8)=3$, where $8$ is the required number of slots to form $(1_14_1)$, and $(17-8)$ the required slots to rotate $1_1$ from node $4$ to node $5$.
As link $(1_12_1)$ needs $8$ time slots to be formed, but $1_1$ rotates counter-clockwise towards node $2$ (opposite direction of final node $5$), the MALT is $20-8-(17+8)=-13$, which is corrected to $0$. Hence, all links can be used except $(1_12_1)$.

\subsubsection{Link ranking phase}

This phase ranks each possible link with a score (line 6), using the following seven attributes:

\begin{itemize}
    \item Number of time slots required to form the link;
    \item Maximum active link time;
    \item Number of initially unused interfaces (if $d_n$ and/or $d'_{n'}$ are not part of a link at $k=1$);
    \item Initial state link $\left((d_nd'_{n'}) \in \mathcal{X}_{init}\right)$;
    \item Final state link $\left((d_nd'_{n'}) \in \mathcal{X}_{end}\right)$;
    \item Traffic demand from both interfaces in $\mathcal{X}_{init}$;
    \item Traffic demand from both interfaces in $\mathcal{X}_{end}$.
\end{itemize}

All attribute scores are normalized and multiplied by a set of weights $\mathcal{W}$, where its elements are $0 \le w_i \le 1, i=1,\dots,7$. Increasing $w_i$ for a given attribute favors links that have a higher value for that attribute (see Section~\ref{sec:evaluation} for tuning these weights).
The sum of all weighted attributes sets the link score (line 7), which is added to the set of link scores (line 8). The final link score list is sorted in decreasing order (line 10).

\subsubsection{Final link selection phase}
The set of final links to form is obtained by extracting links from the sorted ranked link list (line 12) until it is empty (line 11). The extracted links are added to the set of final links (line 13), and both interfaces ($d_n$ and $d'_{n'}$) are flagged as used in the final configuration (line 14). The ranked list is updated after each extraction, removing all remaining links that have $d_n$ or $d'_{n'}$ (line 15).

\subsubsection{Assignment of interface movement decision variables}
The movement decision variables $CW$ and $CCW$, which define the clockwise and counter-clockwise movement over all $k$ slots, are fixed for all involved interfaces. Each link $(d_nd'_{n'})$ from the  final link set (line 17) is  processed in two steps:

\begin{enumerate}
    \item We set $CW$ and $CCW$ for both $d_n$ and $d'_{n'}$ interfaces (line 18). If only one interface needs to rotate, the movement starts at $k=1$. When both interfaces need to rotate,  both interfaces are set to be aligned at the same time slot. With link $(1_14_1)$ from Figure~\ref{fig:greedy-example}, $1_1$ starts its counter-clockwise rotation at $k=1$, while $4_1$ starts moving clockwise at $k=3$. If any of the interfaces need to form a different link from $\mathcal{X}_{end}$, their movement is scheduled after its MALT is reached. For link $(1_14_1)$, the $CW$ values of $1_1$ are then set from $k=8+3$, so it can rotate towards node $5$;

    \item If interface $d_n$ is used in a link from $\mathcal{X}_{end}$ (line 19) with a different interface from a node $d''$ (line 20), we verify if the respective interface $n''$ was not processed during the final link selection phase (line 21). If this is true, $CW$ and $CCW$ are set to  have $d''_{n''}$ aligned with $d$ at $k=K$ (line 22). The same is verified with $d'_{n'}$ in lines 25-30, e.g. for link $(1_14_1)$, we set the $CCW$ values of $5_1$ from $k=20-5$, so $(1_15_1)$ can be formed at $k=K$.
\end{enumerate}

\subsubsection{Post-processing and routing phase}
\label{subsubsec:routing}

After assigning all interface movement decision variables, the BH topology in each time slot is computed by incrementing the $\boldsymbol{A}^0$ values by $\theta \times (CW -CCW)$ (line 32). For each topology, a link is formed when the alignment values from each interface pair $d_n$ and $d'_{n'}$ are the same as $\boldsymbol{V}_{dd'}$ and $\boldsymbol{V}_{d'd}$, if $\boldsymbol{\delta}_{dd'} = 1$. For all $K$ topologies, we then compute the optimal routing using an LP to minimize the total reconfiguration packet loss (line 33), according to $\rho$ and $\boldsymbol{R}$. The LP problem uses three continuous variables that specify the input rate at each node $d \in \mathcal{I}$, the data rate between each node pair $d$ and $d'$, and the loss on each node $d \in \mathcal{D}$. It uses flow conservation constraints, as in Equation 8 from \cite{8647747}, which guarantee that the total input rate and packet loss of each node are equal to their total output rate and traffic demand.

\subsection{Multi-start Randomized Greedy-SBRA}
\label{subsec:ms-greedy}
The best results with the Greedy-SBRA are achieved when an optimal weight set $\mathcal{W}$ is used in its link ranking phase. Yet, finding the optimal weight set with exhaustive parameter tuning can be time-consuming. Instead of running a single Greedy-SBRA iteration with random weights or with a generic weight set, the algorithm can be improved by running a multi-start variant that tests $\Omega$ random weight sets.
For further diversification, the link selection phase is modified by randomly choosing one link from the first $\xi$ elements from the ranked link list (modification of line 12 from Algorithm~\ref{alg:greedy-algorthm}). If $\xi=1$, the approach is pure greedy, while for $\xi>1$ the approach is randomized, returning different results on each run.
For each weight set $\mathcal{W}$, the Greedy-SBRA starts with $\xi=1$, followed by $I$ iterations with $\xi=E$. Consequently, the Greedy-SBRA is executed $\Omega\times(1+I)$ times, and the best-found solution is returned as the final one.

\section{Evaluation}
\label{sec:evaluation}
In this section, we aim to answer the following questions: \textit{how good is the solution quality of our heuristics} and \textit{are they suitable for online reconfiguration?}

\subsection{Baselines}
To compare the solution quality of the greedy algorithms, we benchmark against the following algorithms:

\begin{itemize}
  \item \textbf{Optimal results (MILP):} To calculate the optimal reconfiguration sequence that minimizes packet loss, we use the exact MILP from \cite{8647747}. Because the problem is NP-hard, we can only solve topologies with low $K$ time slots (e.g. up to 21) and low number of nodes $D$. Thus, we can only compare the results for smaller problem instances.
  \item \textbf{PVF-MILP:} To solve larger problem instances in reasonable time, we run a partial variable fixing MILP (PVF-MILP), which finds sub-optimal solutions. It is derived from the MILP in \cite{8647747}, by fixing the movement decision variables for the interfaces from the links in $\mathcal{X}_{end}$, having them reach their final destination as early as possible.
The remaining interfaces' alignment and traffic routing is then optimally solved using the MILP (which has reduced complexity).

 \item \textbf{All links fixed:} This algorithm fixes all movement decision variables and then solves traffic routing using the same LP from Section~\ref{subsubsec:routing}. The movement decision variables from the links in $\mathcal{X}_{end}$ are fixed according to PVF-MILP, while the remaining interfaces are left in their initial state, i.e. no intermediate links are established.
\end{itemize}

\subsection{Experimental Setup}
\label{subsec:SimSetup}

\begin{table}
\begin{center}
\caption{Parametrization of the used topologies}
\label{tab:topologies}	
\begin{tabular}{ c | c | c | c | c | c }
\textbf{Topology}						& \textbf{$D$}			& \textbf{$|\mathcal{I}|$}	& \textbf{$N$}		& \textbf{Users}		& \textbf{$\sum \rho_d$}  		\\
\hline
\multirow{2}{*}{\textit{Grid}}	& \multirow{2}{*}{16}	& \multirow{2}{*}{1}		& 3  				& \multirow{2}{*}{100}	& \multirow{2}{*}{6400 Mbps}	\\
							 			& 						&							& 4					&						&								\\ \hline
\multirow{2}{*}{\textit{Hexagon small}}	& \multirow{2}{*}{19}	& \multirow{2}{*}{1}		& 3					& \multirow{2}{*}{105}	& \multirow{2}{*}{6650 Mbps}	\\
					 					& 						&							& 4					&						&								\\ \hline
\multirow{2}{*}{\textit{Hexagon large}}	& \multirow{2}{*}{37}	& \multirow{2}{*}{2}		& 3					& \multirow{2}{*}{210}	& \multirow{2}{*}{14150 Mbps}	\\
					 					& 						&							& 4					&						&								\\
\hline					 					
\end{tabular}
\end{center}
\end{table}

We evaluate the proposed heuristics using multiple topologies, varying the number of mesh nodes $D$ and network interfaces $N$ (Table~\ref{tab:topologies}). Each topology has a number of core network connected nodes, $|\mathcal{I}|$, and number of served users, with respective demands.
The \textit{Grid} topology is formed by a $4\times4$ mesh where every node is placed over every $s$ =  180 m and then shifted on the $x$ and $y$ axis by two independent random variables following a normal distribution, with $\mu=0$ and $\sigma=\frac{s}{8}$ \cite{8647747}.
The \textit{Hexagon} topologies follow a hexagonal layout, with 19 nodes (small) and 37 nodes (large), spaced with 140 m increments on both $x$ and $y$ axis \cite{CBR}.
For simplicity purposes, we consider $\theta=10^{\circ}$ and $\tau=0.2$ s, corresponding to a 360\textdegree ~rotation in 7.2 s.
The maximum capacity of each mmWave link is computed using \cite{7523521} for the 60 GHz band, with a transmit power level of 23 dBm. We use a truncated Shannon equation to limit the data rate between 4.64 Gbps and 1 Gbps, based on the channel quality \cite{8647747}. 
Each user demand is randomly assigned to a node $d \in \mathcal{D}$ for a given input probability (i.e. 70\% of the users require 50 Mbps, 20\% need 75 Mbps and 10\% have a 100 Mbps demand). The total demand $\sum \rho_d$ is selected to be large enough to congest the links from the nodes $d \in \mathcal{I}$ when $N=3$, and to provide a lower load when $N=4$.
With lower $N$ values, i.e. $N=2$, the BH cannot handle the total demand, resulting in packet loss at each C2 state, which leads to the increase of packet loss with higher $K$ values, even with the optimal reconfiguration \cite{8647747}. Moreover, with $N=5$ or higher, the BH links would be vastly underutilized and the optimal problem complexity significantly increases, not allowing to use the original MILP from \cite{8647747} to benchmark optimal reconfiguration solutions.

The initial C1 and final C2 states are generated by creating two different traffic demands. On each, we compute the optimal routing using a simpler variant of the exact MILP that does not have the interface movement related constraints \cite{8647747}. The demand values from C2 are used to set $\rho_d, \forall d \in \mathcal{D}$. We randomly select values $[0, 360[$ (multiples of $\theta$) to populate $\boldsymbol{A}^0$, for the initially unused interfaces. The selected $K$ values for evaluation are $19$, $20$, $21$, $25$, $30$, and $35$, where $K=19$ is the minimum number of time slots required to reconfigure a topology when at least one interface must rotate 180\textdegree.

All algorithms run in Matlab R2017a, using an Intel Xeon E5-2630 2.30GHz CPU (20 cores) and 184 GB RAM. The MILP is solved using Gurobi 7.5.2 \cite{gurobi}.

\subsection{Link Attribute Weight Selection}
\label{subsec:Parametertuning}
The Greedy-SBRA results depend on the selected weight set $\mathcal{W}$ used to rank potential links. To determine good weights, we ran the Greedy-SBRA with multiple weight combinations for each problem instance. Each attribute weight was varied between values $\{0, 0.33, 0.66, 1.0\}$, generating a total of $4^7$ combinations. The results of running Greedy-SBRA with the best found weights for each topology, $K$, and $N$ are denoted as \textit{Tuned Greedy-SBRA}.
While such offline parameter tuning yields good results and is more scalable than the MILP, it is not suitable for online usage, as it requires new tuning every time the input problem parameters change, e.g. $K$, $N$, or $\rho_d$.

In addition, since establishing a single generic weight profile to be used for all instances did not yield satisfactory results, we propose running the multi-start algorithm variant from Section~\ref{subsec:ms-greedy}, referred to as \textit{MS Greedy-SBRA}. The results are shown for 20 random weight sets ($\Omega=20$). For each weight set $\mathcal{W}$, the link selection phase was run once as pure greedy ($\xi=1$) and  10 times ($I=10$) with $E=10$, randomly choosing one of the best 10 links from the ranked link list in each step. Hence, the \textit{MS Greedy-SBRA} ran 220 iterations of the Greedy-SBRA. Running more iterations can slightly improve the solution quality at the expense of higher runtime.

\subsection{Numerical Results}
\label{subsec:results}

\begin{figure}[t]%
\centering
\begin{subfigure}{.49\columnwidth}
\includegraphics[width=\columnwidth]{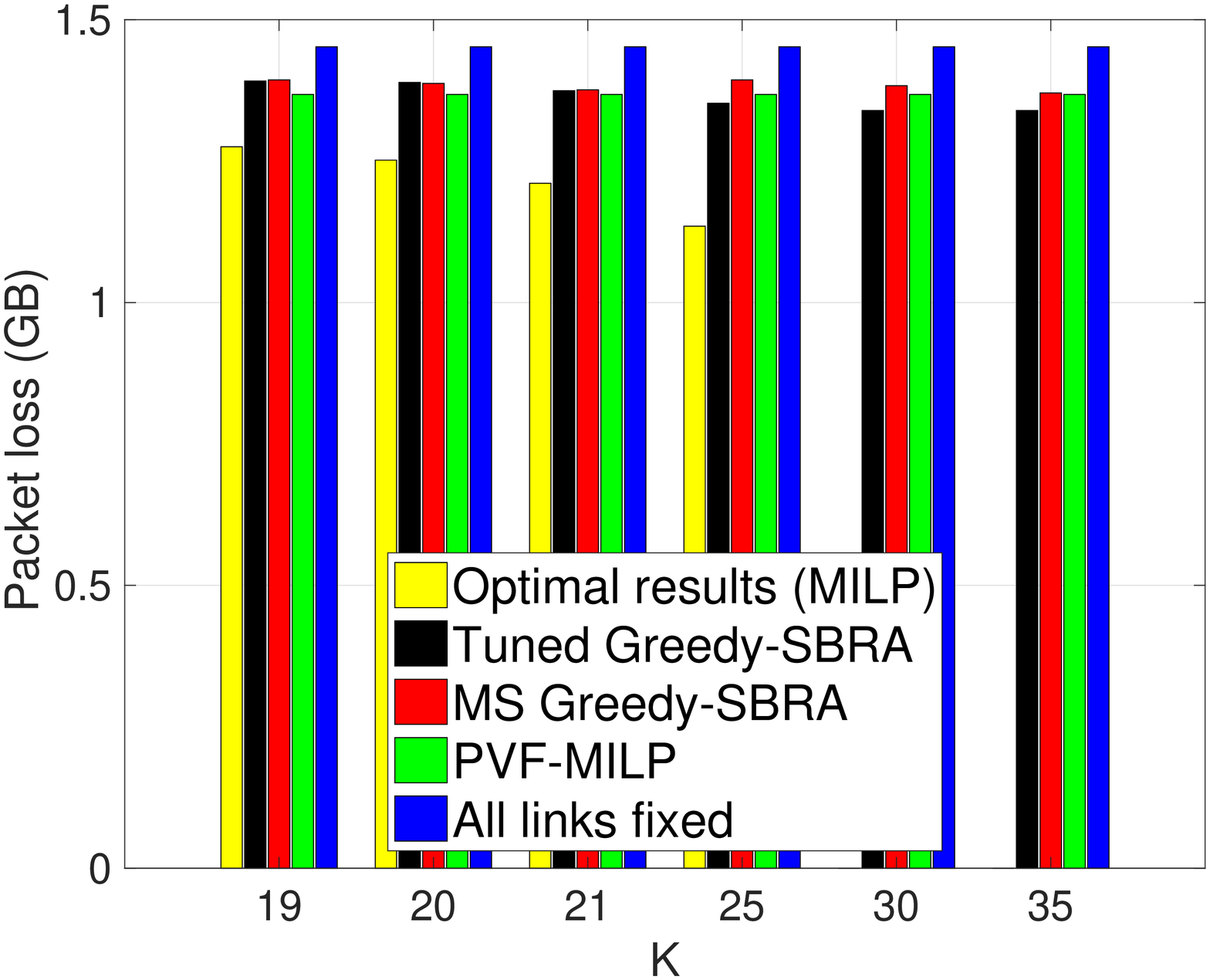}%
\caption{\textit{Grid} with $N=3$}%
\label{fig:shift-d16-n3}%
\end{subfigure}\hfill%
\begin{subfigure}{.49\columnwidth}
\includegraphics[width=\columnwidth]{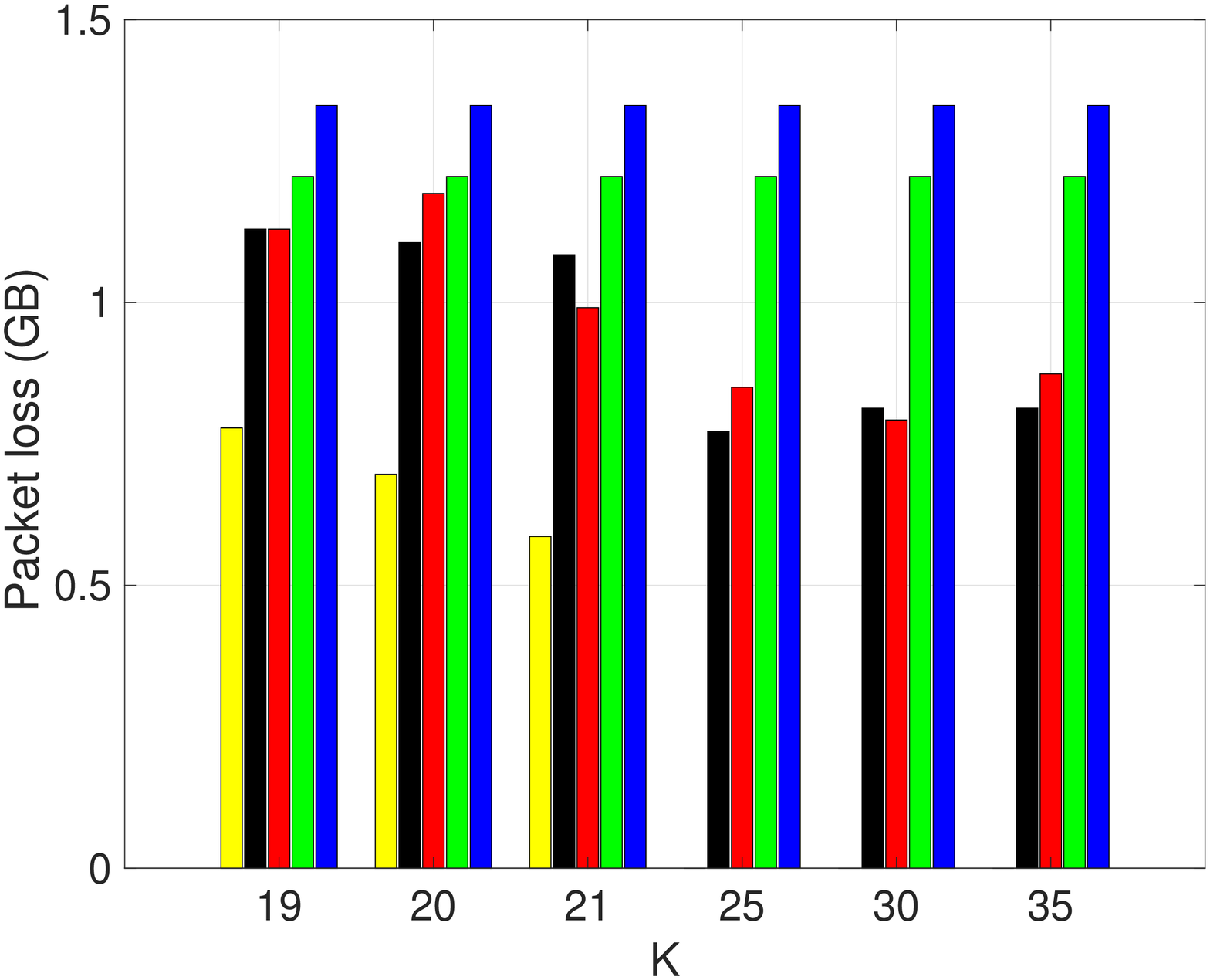}%
\caption{\textit{Grid} with $N=4$}%
\label{fig:shift-d16-n4}%
\end{subfigure} % \hfill %\\ %
\caption{Total packet loss versus number of $K$ time slots for the \textit{Grid} topology.}
\label{fig:shift-d16-loss}
\end{figure}

\begin{figure}[t]%
\centering
\begin{subfigure}{.49\columnwidth}
\includegraphics[width=\columnwidth]{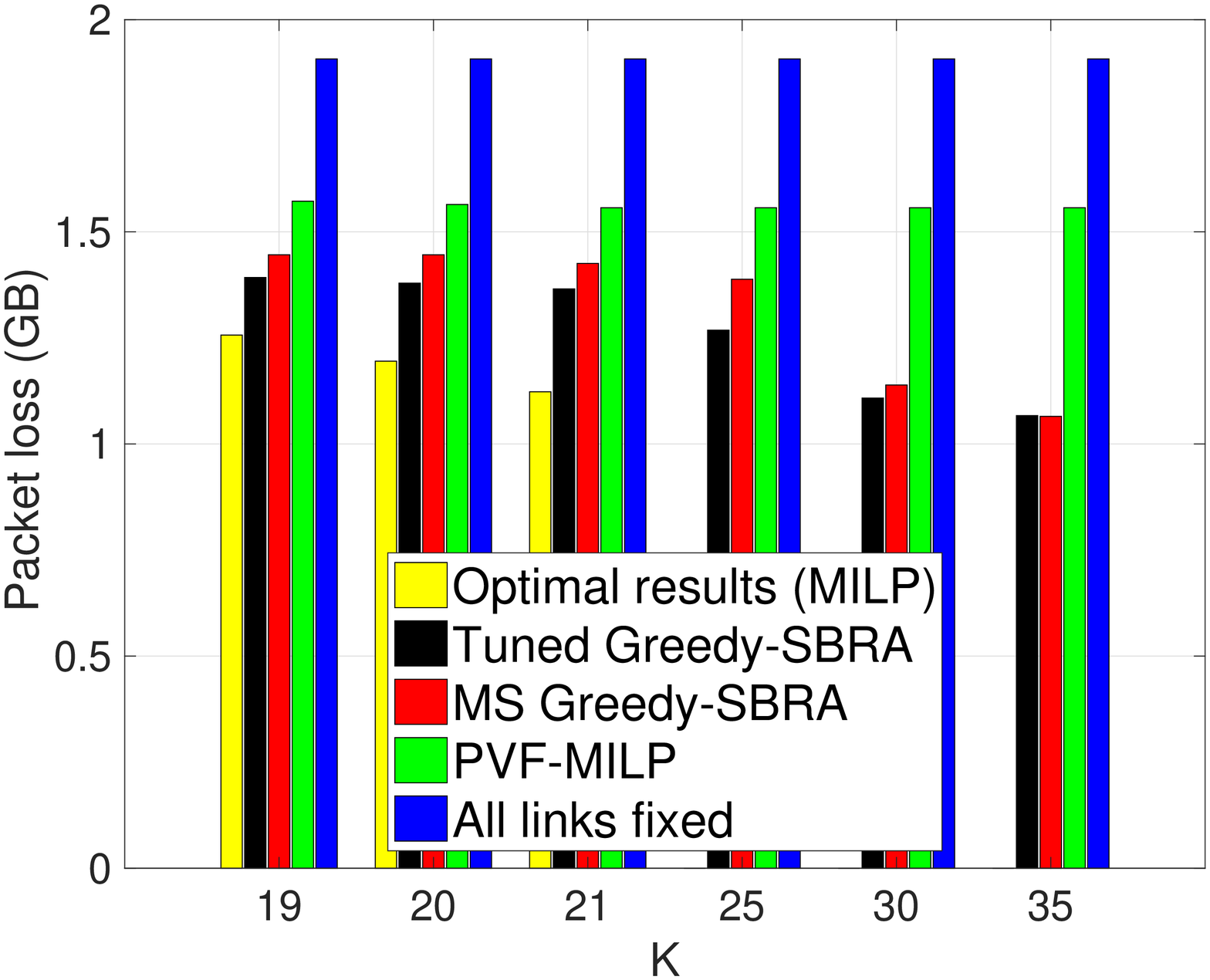}%
\caption{\textit{Hexagon small} with $N=3$}%
\label{fig:hex-d19-n3}%
\end{subfigure} \hfill%
\begin{subfigure}{.49\columnwidth}
\includegraphics[width=\columnwidth]{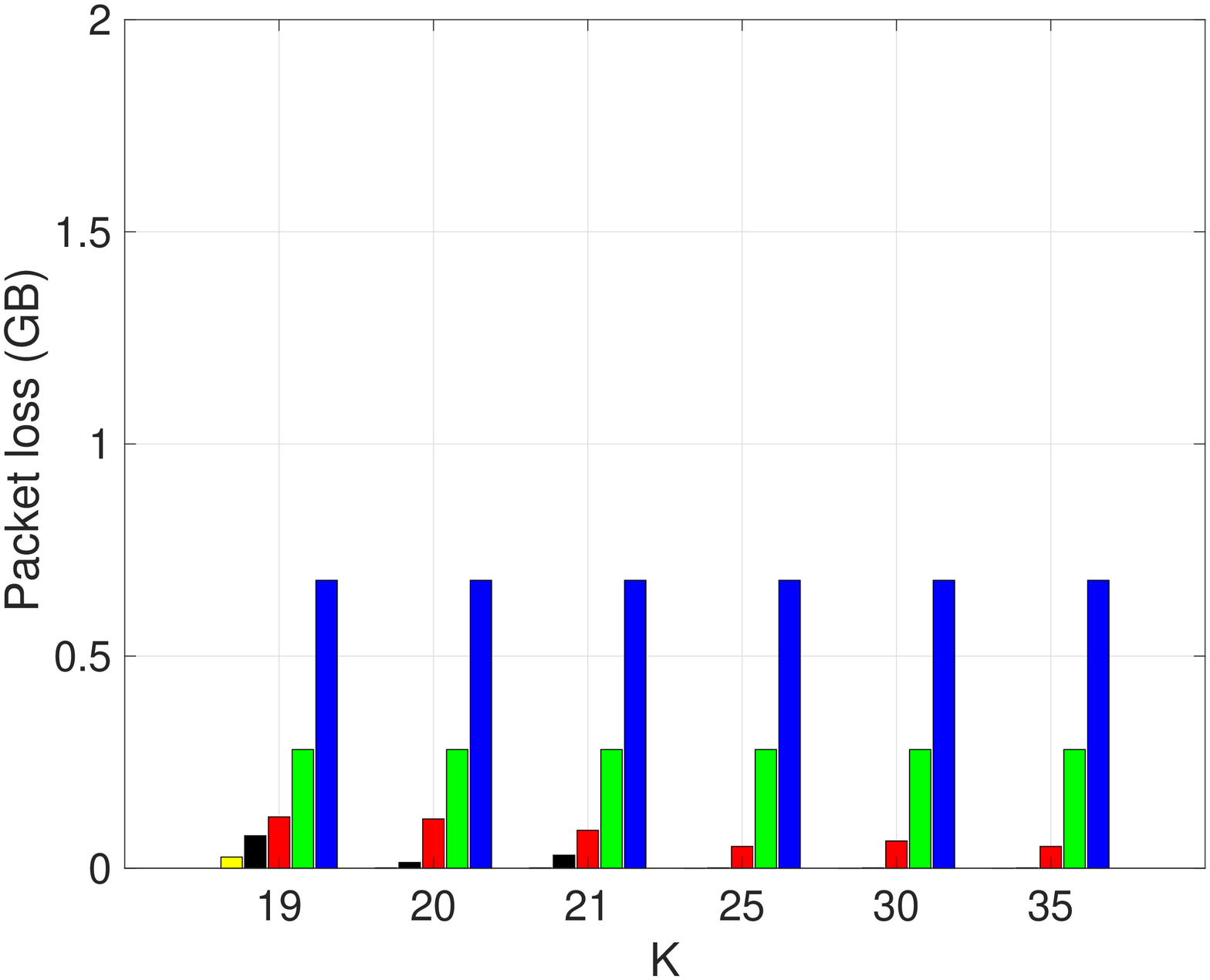}%
\caption{\textit{Hexagon small} with $N=4$}%
\label{fig:hex-d19-n4}%
\end{subfigure} %\hfill%
\caption{Total packet loss versus number of $K$ time slots for the \textit{Hexagon small} topology.}
\label{fig:hex-d19-loss}
\end{figure}

\begin{figure}[t]%
\centering
\begin{subfigure}{.49\columnwidth}
\includegraphics[width=\columnwidth]{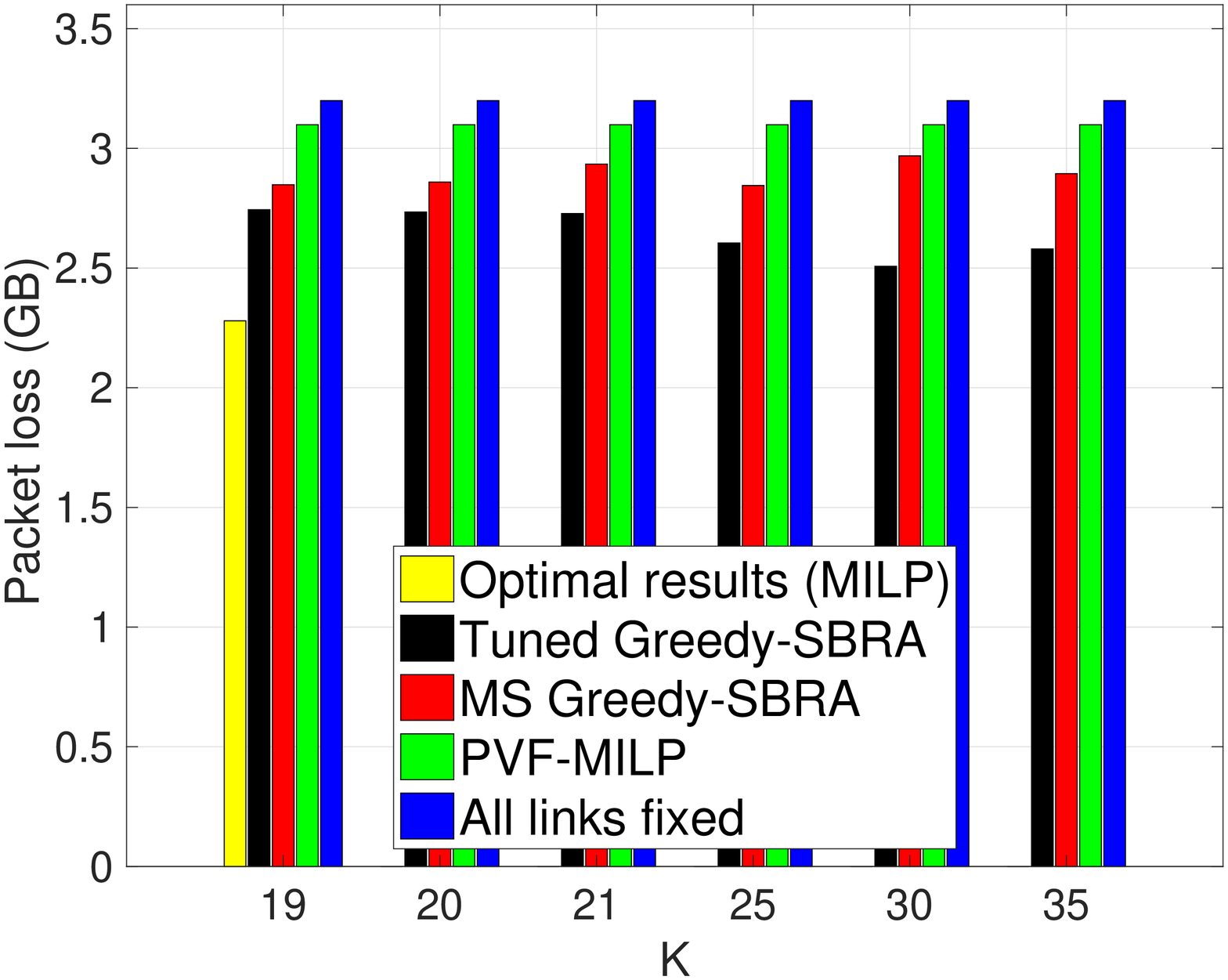}%
\caption{\textit{Hexagon large} with $N=3$}%
\label{fig:hex-d37-n3}%
\end{subfigure} \hfill%
\begin{subfigure}{.49\columnwidth}
\includegraphics[width=\columnwidth]{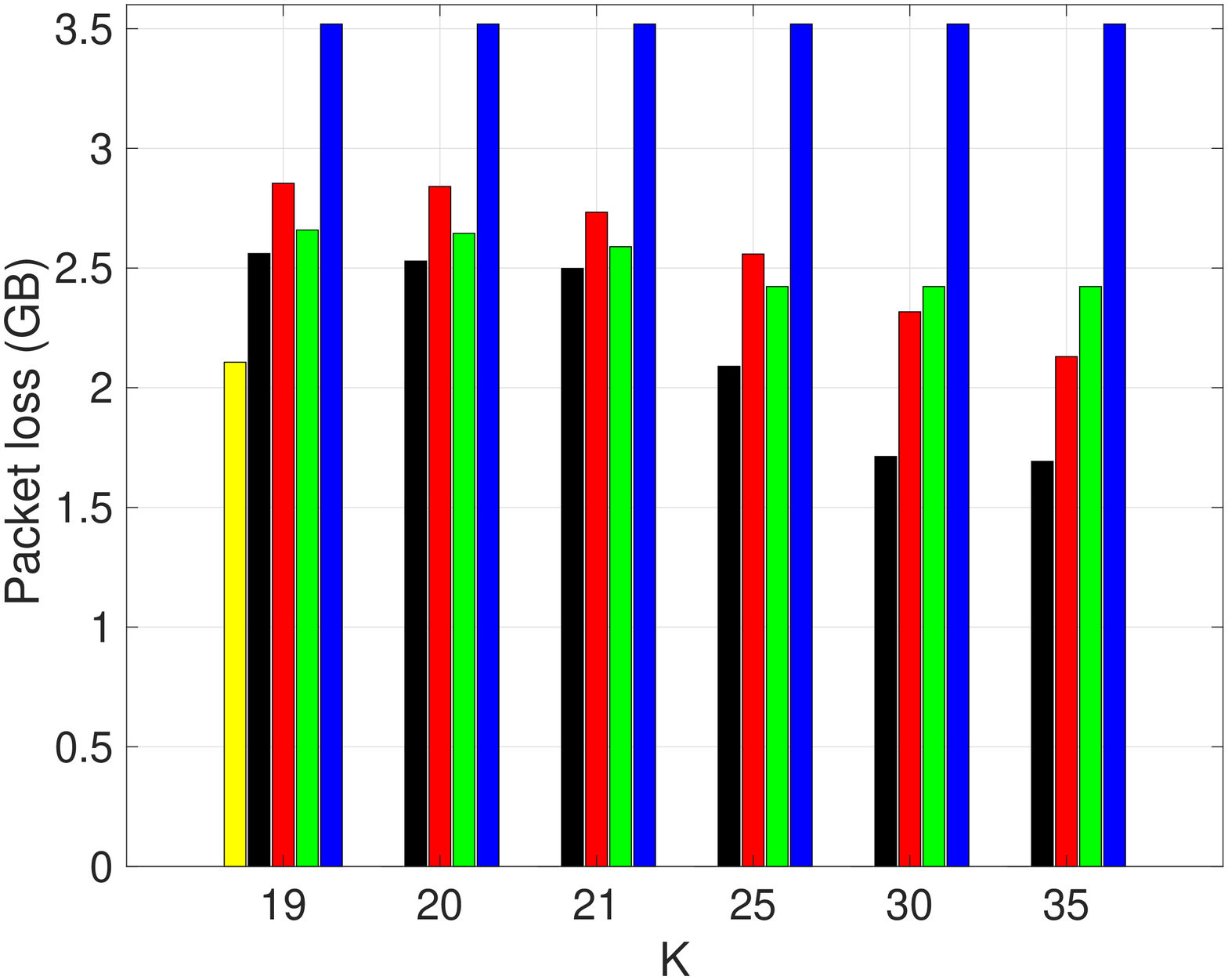}%
\caption{\textit{Hexagon large} with $N=4$}%
\label{fig:hex-d37-n4}%
\end{subfigure}
\caption{Total packet loss versus number of $K$ time slots for the \textit{Hexagon large} topology.}
\label{fig:hex-d37-loss}
\end{figure}

Figure~\ref{fig:shift-d16-loss} shows packet loss for the \textit{Grid} topology for $N=3$ and $N=4$, for all the described algorithms. 
We present the total packet loss instead of the packet loss rate, since the total loss rate decreases substantially with higher $K$ values (caused by more data being transferred during the reconfiguration, with the packet loss not increasing with the optimal reconfiguration solution \cite{8647747}). Therefore, we are mostly interested in the overall packet loss during the reconfiguration, when varying the used $K$ values.
Because of its complexity, the MILP could only solve problems up to $K=25$ with $N=3$ and up to $K=21$ with $N=4$. The optimal packet loss decreases when $K$ is increased, as a larger reconfiguration interval allows more backup paths to be established, while reducing the probability that  the rotation of multiple interfaces from the same node overlaps in time (e.g. if all interfaces from a gateway node are rotating, all remaining nodes are disconnected).
The \textit{All links fixed} algorithm has the same loss for all $K$ values (1.45 GB, 14\% more loss than the optimal solution with $K=19$), as its behavior does not change when varying $K$. 
The \textit{PVF-MILP} results are also constant with $K$ for this topology (7\% more loss compared to the optimal solution for $K=19$), since most of its interfaces are used in the final state  and its nodes do not have a high number of possible neighbors. Thus, the optimization of the unused interfaces' movement does not vastly improve the solution.
Both greedy algorithms achieve similar performance, although \textit{Tuned Greedy-SBRA} has slightly lower loss than \textit{MS Greedy-SBRA}.
The results of both algorithms are close to the ones from \textit{PVF-MILP} (0.2\% more loss) and better than \textit{All links fixed} by 5\% on average.
With $N=4$ (Figure~\ref{fig:shift-d16-n4}), we observe a $\approx$ 39\% decrease of packet loss from the optimal solution with $K=19$, when compared to $N=3$. Moreover, the packet loss decreases by 25\% from $K=19$ to $K=21$. When the number of antennas per node increases, the number of possible links during the reconfiguration also increases.
The total loss with the \textit{All links fixed} and \textit{PVF-MILP} algorithms also decreased to 1.35 GB and 1.22 GB. Yet, we observe a large gap against the optimal solution at higher $K$ values.
The packet loss obtained by \textit{Tuned Greedy-SBRA} and \textit{MS Greedy-SBRA} outperform the results obtained by \textit{All links fixed} and \textit{PVF-MILP} in all cases, generally decreasing when increasing $K$. For example, with the \textit{Tuned Greedy-SBRA}, there is 32\% less packet loss at $K=25$, when compared to $K=19$. 
However, please note that while the optimal packet loss never increases with a higher $K$, this is not always the case with the Greedy-SBRA. Namely, as each link is configured to remain active during its MALT, increasing $K$ can lead to reconfiguration intervals where the BH stays in an intermediate topology with high packet loss, for a longer period of time. When that happens, a higher $K$ can lead to more packet loss.

With the \textit{Hexagon small} topology, it is possible to solve the MILP for values up to $K=21$, for both $N=3$ and $N=4$ (see Figures~\ref{fig:hex-d19-n3} and \ref{fig:hex-d19-n4}). With $N=3$, the \textit{All links fixed} shows almost 52\% more packet loss compared to the optimal solution with $K=19$. With \textit{PVF-MILP}, the total loss decreases to $\approx$ 1.6 GB (25\% more than the optimum). For this topology, the \textit{Tuned Greedy-SBRA} slightly outperforms \textit{MS Greedy-SBRA} in most cases, but both algorithms outperform the \textit{All links fixed} and \textit{PVF-MILP}.
With the total loss decreasing with the increase of $K$ in this topology, the results from the \textit{Tuned Greedy-SBRA} are lower than the best optimal solution found, when we allow more reconfiguration time slots ($K \geq 30$). 
This shows that using a fast-heuristic to solve the BH reconfiguration that allows higher $K$ values can lead to less packet loss, when compared to the optimal solution found with a MILP for solvable $K$ values.
With $N=4$ in this topology, the MILP reconfigures the BH at negligible loss (1 MB) for $K=20$ and $K=21$. 
Yet, the \textit{PVF-MILP} and \textit{All links fixed} algorithms result in large packet loss.
The proposed greedy approaches approximate the optimum, and the packet loss decreases with higher $K$ values.

Figure~\ref{fig:hex-d37-loss} shows the loss for the \textit{Hexagon large} topology, where the MILP can only solve for $K=19$. For this topology, the optimal solution only improves 7\% when increasing $N$.
For $N=3$, both greedy algorithms outperform the variable fixing heuristics. The \textit{Tuned Greedy-SBRA} gives the best heuristic results, with a total loss 20\% higher than the optimal solution with $K=19$, and slightly decreasing with higher $K$ values.
For $N=4$, the \textit{PVF-MILP} outperforms the \textit{MS Greedy-SBRA} for $K \leq 25$, while the \textit{Tuned Greedy-SBRA} still gives the best results among the heuristics. Note, that for $K \geq 25$, the \textit{Tuned Greedy-SBRA} achieves results up to 20\% lower than the optimal solution obtained with the MILP, for $K=19$.

\begin{table}
\begin{center}
\caption {Execution time for all algorithms (in seconds), with all tested topologies, for $K=19$, $25$, and $35$.}
\label{table:algorithm-time}	
\resizebox{\columnwidth}{!}{ 
\begin{tabular}{ cccccccc }
  \multirow{2}{*}{\textbf{\textit{K}}}	&
  \multirow{2}{*}{\textbf{Algorithm}}	&
  \multicolumn{2}{c}{\textbf{\textit{Grid}}}	&
  \multicolumn{2}{c}{\textbf{\textit{Hexagon small}}}	&
   \multicolumn{2}{c}{\textbf{\textit{Hexagon large}}} \\
	& 	& \textbf{\textit{N=3}}	& \textbf{\textit{N=4}}	& \textbf{\textit{N=3}}	& \textbf{\textit{N=4}}	& \textbf{\textit{N=3}}	& \textbf{\textit{N=4}} \\	\hline			
\multirow{5}{*}{$19$}			& MILP					& 115.76	& 144.56	& 1233.64	& 887.23	& 1647114.47	& 39841.74 \\
								& Tuned Greedy-SBRA	    & 0.33		& 0.10		& 0.12		& 0.15		& 0.43			& 0.50 \\
								& MS Greedy-SBRA		& 16.63		& 14.86 	& 17.96 	& 22.89 	& 99.31			& 99.88 \\
								& PVF-MILP				& 18.00		& 28.76		& 30.98		& 516.32	& 60.55			& 449.86 \\
								& All links fixed		& 0.07		& 0.05		& 0.07		& 0.06		& 0.33			& 0.32 \\ \hline
\multirow{5}{*}{$25$}			& MILP					& 25859.72	& ---		& ---		& ---		& ---			& --- \\
								& Tuned Greedy-SBRA	    & 0.12		& 0.13		& 0.17		& 0.20		& 0.63			& 0.71 \\
								& MS Greedy-SBRA		& 19.23 	& 16.81		& 33.57		& 30.13		& 154.09		& 150.83 \\
								& PVF-MILP				& 53.40		& 95.25		& 90.80		& 1099.79	& 127.41		& 1607.09 \\
								& All links fixed		& 0.07		& 0.08		& 0.11		& 0.11		& 0.51			& 0.51 \\ \hline
\multirow{5}{*}{$35$}			& MILP					& ---   	& ---		& ---		& ---		& ---		    & --- \\
                                & Tuned Greedy-SBRA	    & 0.18		& 0.19		& 0.24		& 0.27		& 1.03			& 1.14 \\
								& MS Greedy-SBRA		& 37.08		& 37.86		& 51.20		& 52.55		& 247.96		& 246.97 \\
								& PVF-MILP				& 289.11	& 871.05	& 250.39	& 1059.70	& 1036.83		& 36015.85 \\
								& All links fixed		& 0.12		& 0.12		& 0.17		& 0.17		& 0.89			& 0.89 \\ \hline
\end{tabular}
}
\end{center}
\end{table}

Table~\ref{table:algorithm-time} shows the execution time for all algorithms for $K=19$, $25$, and $35$. The solver times for the optimal MILP exponentially increase with the problem size (therefore, not all problem instances could be solved). The runtime of the \textit{PVF-MILP} also increases dramatically when increasing $N$ from $3$ to $4$, although it was able to solve all problem instances. However, for large topologies, this approach is also not suitable for online optimization. The \textit{All links fixed} algorithm has overall the lowest running times, since all movement variables are fixed without complex pre-processing. 
The \textit{Tuned Greedy-SBRA} runs faster than 1 s for most cases, but the shown execution times do not include the time needed for parameter tuning, which was in the order of hours.
The \textit{MS Greedy-SBRA} runs 220 iterations of the Greedy-SBRA with different weights, yielding somewhat higher execution times, but without prior offline tuning.
As future work, the runtime of the \textit{MS Greedy-SBRA} can be significantly reduced by parallelizing its execution on multiple CPU cores (in this work, we used a single core). In addition, we can dynamically adapt its $\Omega$ and $I$ values based on each iteration's runtime and on the improvement over the previously found best solution.

\section{Conclusion}
\label{sec:conclusions}
In this paper, we propose greedy-based heuristic algorithms that orchestrate the reconfiguration of a small cell wireless BH. The BH radios use directional mmWave antennas, which are aligned through steerable mechanical devices, that need to rotate to form new links. The reconfiguration should be seamless to existing traffic and involves the rotation of the mmWave interfaces, link establishment, and BH routing updates.
In our evaluation, we compare the performance of our heuristics against different benchmarks in terms of packet loss and execution time, for different topologies and scenarios. 
Our results show that our algorithms run fast, can be used for the online optimization of BH reconfiguration, while giving good quality solutions with respect to packet loss.
As future work, we will improve the parameter tuning by using machine learning approaches and integrate our algorithms with a real SDN-based small cell mesh BH testbed.

\bibliographystyle{IEEEtran}
\bibliography{gc2019}

\end{document}